\documentclass[aps,twocolumn,final]{revtex4}%
\usepackage{amssymb}
\usepackage{amsfonts}
\usepackage{amsmath}
\usepackage{graphicx}%
\begin{document}
\title{How quantum bound states bounce and the structure it reveals}
\author{{Dean~Lee, Michelle Pine}}
\affiliation{Department of Physics, North Carolina State University, \linebreak Raleigh, NC
27695, USA}

\begin{abstract}
We investigate how quantum bound states bounce from a hard surface. \ Our
analysis has applications to \textit{ab initio} calculations of nuclear 
structure and
elastic deformation, energy levels of excitons in semiconductor quantum dots
and wells, and cold atomic few-body systems on optical lattices with sharp
boundaries. \ We develop the general theory of elastic reflection for a
composite body from a hard wall. \ On the numerical side we present \textit{ab
initio} calculations for the compression of alpha particles and universal
results for two-body states. \ On the analytical side we derive a universal
effective potential that gives the reflection scattering length for shallow
two-body states.

\end{abstract}

\maketitle

In this letter we consider the elastic scattering of quantum bound states from
a hard surface. \ Elastic scattering from a hard surface provides direct
information about the structure of the bound state and how it behaves under
spatial confinement and compression. \ Spatially-confined bound states are
interesting quantum systems with important practical applications. \ For
example in quantum dots and wells, electron-hole pairs are confined to the
interior of nanoscale semiconductor structures. \ The size and shape of the
structure strongly influences the density of exciton states, thereby allowing
control of current tunneling as well as photon absorption and emission. \ See
for example Ref.~\cite{Takagahara:1992a,Marzin:1994a,Pfalz:2005a,He:2005a}.
\ In our analysis we present universal results for direct band-gap
semiconductors, relating exciton energies to the binding energy at infinite
volume, the effective masses of the electrons and holes, and the geometry of
the nanostructure.

Confined systems can also be produced in cold atomic experiments. \ A quantum
well can be constructed using an optical lattice with additional lasers
focused to produce sharp repulsive boundaries. \ Similar ideas have been
proposed for quantum billiards systems \cite{Montangero:2009}. Since a flat
hard-wall boundary has no dimensionful scale, reflection from the wall probes
the universal physics of few-body systems at large scattering length. \ For
example the dimer-wall reflection scattering length for a shallow dimer is
directly proportional to the two-particle scattering length. \ In the
following we will calculate the universal constant of proportionality between
these two scattering lengths.

We consider a non-relativistic bound state in $d$ spatial dimensions with mass
$M$. \ The number of constituent particles comprising the bound state is
arbitrary. \ The bound state scatters elastically against a flat hard-wall
boundary. \ Let $X$ be the distance from the wall to the center of mass of the
bound state. \ We choose the inertial frame where all components of momentum
parallel to the wall surface are zero and construct a standing wave solution
with momenta $\pm p$ perpendicular to the wall. \ We denote the reflection
phase shift as $\delta(p)$, and define the reflection radius as $R(p)=-\delta
(p)/p$. \ This corresponds with the distance between the wall and the closest
node of the asymptotic standing wave, $\Psi_{p}(X)\propto\sin\left[
pX+\delta(p)\right]  $.

The reflection scattering process is analogous to $S$-wave scattering in three
dimensions. \ We can therefore use the effective range expansion,%
\begin{equation}
p\cot\delta(p)=-\frac{1}{a_{R}}+\frac{1}{2}r_{R}p^{2}-\mathcal{P}_{R}%
p^{4}+\cdots, \label{ERE}%
\end{equation}
where $a_{R}$ is the scattering length, $r_{R}$ is the effective range, and
$\mathcal{P}_{R}$ is the shape parameter. \ We note that $a_{R}$ equals the
reflection radius at threshold, $a_{R}=\lim_{p\rightarrow0}R(p)$. \ If the
bound state structure is completely rigid then $R(p)=a_{R}$ for all
$p$.\ \ For realistic systems, however, compression of the bound state
increases with collision energy. \ Hence $R(p)$ decreases monotonically with
$p$, and the rate of decrease indicates the compressibility of the bound state
under unilateral stress. \ The reflection radius is directly related to the
energy of the bound state under spatial confinement. \ For a $d$-dimensional
cube with length $L$ and vanishing Dirichlet boundaries, the energy of the
lowest confined mode is%
\begin{equation}
E(L)=\frac{p^{2}(L)\cdot d}{2M}+O\left(  L^{-4}\right)  ,\quad p(L)=\frac{\pi
}{L-2R\left[  p(L)\right]  }. \label{E_L}%
\end{equation}
This formula arises from the sine function profile with half-wavelength
$L-2R\left[  p(L)\right]  $ in each dimension. \ It can be used to determine
the reflection radius as a function of momentum. \ The $O\left(
L^{-4}\right)  $ error estimate contains corrections due to double-wall
collision effects near the wall intersections.

It is not possible to construct a hard surface for protons and neutrons in
nuclear physics experiments. \ Therefore it may seem that the current
discussion has no direct connection with nuclear physics. \ A similar critique
could be made of L\"{u}scher's analysis of periodic boundaries in finite cubic
volumes \cite{Luscher:1985dn,Luscher:1986pf}. \ However L\"{u}scher's analysis
now provides the theoretical framework for numerous calculations in lattice
quantum chromodynamics \cite{Beane:2005rj,Beane:2006mx,Bernard:2008ax}. \ The
point is that \textit{ab initio} numerical calculations play an increasingly
important role in nuclear theory. \ In some cases the numerical calculations
investigate phenomena directly observed in experiments. \ In other cases the
calculations probe new physics not accessible in experiments. \ Similar to
temperature and chemical potential, different spatial boundaries provide
tunable control parameters for the numerical calculations.

We now briefly discuss one example from nuclear physics. \ There has been much
interest in alpha-particle clusters inside nuclei such as carbon-12
\cite{Tohsaki:2001an,Chernykh:2007a,Suzuki:2007wi}. \ Very recently there have
been \textit{ab initio} lattice calculations of the low-lying states of
carbon-12 using effective field theory \cite{Epelbaum:2011md}. \ In addition
to the ground state and excited spin-2 state, these calculations find for the
first time the Hoyle resonance responsible for carbon formation at stellar
temperatures. \ The lattice data suggests the presence of correlated alpha
clusters. \ Furthermore it appears that the alpha clusters themselves are
compressed in size when bound in nuclei. \ This would be interesting since
there are no low-energy resonances of the alpha particle and little
experimental data on the compression of alpha particles.

We have calculated the energy of an alpha particle at leading order in chiral
effective field theory when confined to a cubic box with vanishing Dirichlet
boundaries. \ We use the same lattice action, algorithms, and codes as in
Ref.~\cite{Epelbaum:2011md}. \ The boundaries are imposed by adding an
essentially infinite potential energy at the wall boundaries. \ We note that
the ultraviolet divergences are independent of long-distance boundary
conditions, and so no new renormalization counterterms are needed. \ We only
summarize the results here, and the numerical details of this calculation will
be discussed in a forthcoming publication \cite{Lee:2011z}.

At leading order we have calculated the energy for cubic boxes with lengths
$L=11.8$~fm, $9.9$~fm, and $7.9$~fm. \ The results are shown in Table
\ref{alpha}. \ The error bars in Table \ref{alpha} are one standard deviation
estimates which include both Monte Carlo statistical errors and uncertainties
due to extrapolation at large Euclidean time. \ The reflection radii in Table
\ref{alpha} can be compared with the root-mean-square (RMS) matter radius of
the alpha particle. \ At leading order we find the RMS matter radius for
pointlike nucleons to be $1.53(4)$~fm.

\begin{table}[tb]
\caption{Momenta and reflection radii for an alpha particle confined to a cube
of length $L$.}%
\label{alpha}
$%
\begin{tabular}
[c]{|c|c|c|}%
$L$ & $p(L)$ & $R[p(L)]$\\\hline\hline
$11.8$ fm & $81(9)$ MeV & $2.1(4)$ fm\\
$9.9$ fm & $97(10)$ MeV & $1.6(3)$ fm\\
$7.9$ fm & $118(10)$ MeV & $1.3(2)$ fm
\end{tabular}
\ \ \ \ \ $\end{table}At low momenta the reflection radius of the alpha
particle is somewhat larger than the RMS matter radius. \ However we see a
substantial decrease of the alpha particle reflection radius with increasing
momentum. \ This data suggests that the alpha particles do compress quite
readily with pressure. \ This appears consistent with the observation that
alpha clusters are compressed in size when bound in nuclei. \ Much more
numerical work is planned to study alpha particles and other nuclei under
confinement pressure.

We now focus on the case of two-body bound states. \ The masses of the two
constituent particles will be denoted $m_{1}$ and $m_{2}$, and $\mu$ is the
reduced mass, $\mu=(m_{1}^{-1}+m_{2}^{-1})^{-1}$. $\ $We assume the two
constituent particles are distinguishable and so the quantum statistics and
spin are irrelevant. \ For the case of electron-hole pairs, $m_{1}$ and
$m_{2}$ correspond with the effective masses of the electrons and holes around
zero-momentum minima. \ Our analysis applies to the case of Wannier excitons
where the exciton size is significantly larger than the Coulomb screening length.

Let $E_{B}$ be the energy of the dimer at infinite volume, and $\kappa_{B}$ be
the binding momentum defined by the relation $E_{B}=-\kappa_{B}^{2}/(2\mu)$.
\ We are assuming that the bound dimer has zero orbital angular momentum and
let $a_{B}$ be the $S$-wave scattering length for the two constituent
particles. \ The total spin of the bound state is irrelevant to our analysis
so long as there is negligible partial-wave mixing. \ We consider the
shallow-binding limit where $a_{B}$ is much larger than the range of the
interaction. \ In this limit $\kappa_{B}=a_{B}^{-1}$, and the low-energy
physics of shallow dimers is independent of the short-distance details of the
interaction. \ It follows that the dimer-wall reflection phase shift is a
universal function of the dimensionless ratio $p/\kappa_{B}$. \ In the
following discussion we present the form of this universal function.

We have calculated the dimer-wall reflection phase shift for the one- and
three-dimensional systems numerically using a Hamiltonian lattice formalism.
\ More details on this lattice formalism can be found in
Ref.~\cite{Lee:2008fa}. \ The two-particle interaction is implemented as an
attractive single-site interaction. \ Two parallel hard-wall boundaries are
placed a distance $L$ apart, and we set the momentum of the dimer parallel to
the wall equal to zero. \ The confinement energy is measured as a function of
$L$ using sparse-matrix eigenvector methods to determine the reflection phase
shift. \ Calculations are repeated several times using successively smaller
lattice spacings to extrapolate to the continuum limit and determine universal
results in the shallow-binding limit. \ For the three-dimensional case we also
perform an infinite-size extrapolation in the dimensions perpendicular to the
wall. \ A large number of separate calculations are required to perform these
extrapolations, and the details of each calculation are described in a
forthcoming publication \cite{Lee:2011z}.

Fig.~\ref{latt_1D} shows the one-dimensional results for the reflection radius
versus dimer kinetic energy, $E_{K}$, in the shallow-binding limit. \ The
results are presented in dimensionless combinations, $\kappa_{B}R(E_{K})$
versus $E_{K}/\left\vert E_{B}\right\vert $. \ We show results for several
different mass ratios, $m_{2}/m_{1}$. \ As expected the reflection radius
decreases monotonically with energy.\ \ One interesting feature is the
dependence on the mass ratio. \ For large $m_{2}/m_{1}$ the reflection radius
is rather large at small energies while significantly decreasing with energy.
\ We explain this mass ratio dependence later in our discussion.

For the case $m_{2}/m_{1}=1$, the one-dimensional problem is integrable and
can be solved exactly using the Bethe Ansatz. \ From the Bethe Ansatz we get%
\begin{equation}
p\cot\delta(p)=-2\kappa_{B}\text{,} \label{ERE_BA}%
\end{equation}%
\begin{equation}
\kappa_{B}R(E_{K})=\frac{1}{2}\sqrt{\frac{\left\vert E_{B}\right\vert }{E_{K}%
}}\tan^{-1}\sqrt{\frac{E_{K}}{\left\vert E_{B}\right\vert }}. \label{R_BA}%
\end{equation}
{
\begin{figure}[ptb]%
\centering
\includegraphics[
height=2.2278in,
width=2.2675in
]%
{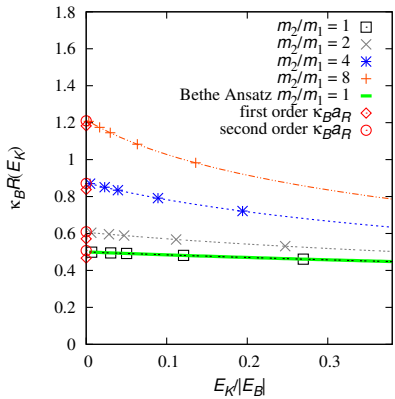}%
\caption{One-dimensional lattice results for the reflection radius versus
dimer kinetic energy in the shallow-binding limit. Also shown are Bethe Ansatz
results for $m_{2}/m_{1}=1$ and first- and second-order results for the
expansion of $\kappa_{B}a_{R}$ described later in the text.}%
\label{latt_1D}%
\end{figure}
As shown in }Fig.~\ref{latt_1D}, the agreement between the lattice
calculations and Bethe Ansatz solution is very good. \ Fig.~\ref{latt_1D} also
shows the first- and second-order analytic results for the expansion of
$\kappa_{B}a_{R}$ described later in our discussion.

The coefficients of the effective range expansion for one-dimensional
dimer-wall scattering are given in Table~\ref{1D_table}. \ The error estimates
are from least-squares fitting used in the lattice extrapolation and the
effective range expansion. \ As one can see from Eq.~(\ref{ERE_BA}), the Bethe
Ansatz gives $\kappa_{B}a_{R}=1/2$ for $m_{2}/m_{1}=1$, with all other higher
coefficients equal to zero. \ This is in full agreement with the lattice
results. \ The results are in principle universal and can be checked using
other theoretical methods or perhaps experiments such as cold atomic dimers on
a one-dimension optical lattice with sharp boundaries.\begin{table}[tb]
\caption{Coefficients of the effective range expansion for one-dimensional
dimer-wall scattering.}%
\label{1D_table}
$%
\begin{tabular}
[c]{|c|c|c|c|}%
$m_{2}/m_{1}$ & $\kappa_{B}a_{R}$ & $\kappa_{B}r_{R}$ & $\kappa_{B}^{3}P_{R}%
$\\\hline\hline
1 & $0.4999(2)$ & $0.005(7)$ & $0.002(3)$\\
2 & $0.6065(2)$ & $-0.074(2)$ & $-0.006(2)$\\
4 & $0.8747(2)$ & $0.115(2)$ & $0.006(2)$\\
8 & $1.2149(2)$ & $0.460(2)$ & $0.008(2)$%
\end{tabular}
\ \ \ \ \ $\end{table}

\begin{figure}[ptb]%
\centering
\includegraphics[
height=2.2278in,
width=2.2675in
]%
{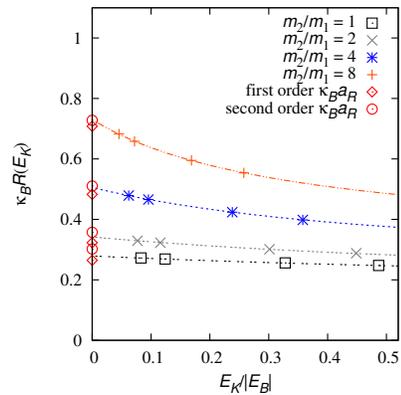}%
\caption{Three-dimensional lattice results for the reflection radius versus
dimer kinetic energy in the shallow-binding limit. \ Also shown are first- and
second-order results for $\kappa_{B}a_{R}$ described later in the text.}%
\label{latt_3D}%
\end{figure}

Fig.~\ref{latt_3D} shows the reflection radius versus dimer kinetic energy for
the three-dimensional system in the shallow-binding limit. \ \ Again the
results are presented in dimensionless combinations, $\kappa_{B}R(E_{K})$
versus $E_{K}/\left\vert E_{B}\right\vert $. \ The reflection radius is nearly
a factor of two smaller than in the one-dimensional case. \ This is reasonable
considering the difference between compression of a one-dimensional object
versus compression of a three-dimensional sphere along one direction. \ As in
the one-dimensional case, we see the same dependence on the mass ratio. $\ $At
large $m_{2}/m_{1}$ the reflection radius is large at small energies while
becoming substantially smaller with increasing energy. \ Fig.~\ref{latt_3D}
also shows the first- and second-order results for the asymptotic expansion of
$\kappa_{B}a_{R}$ described later in our discussion.%

The coefficients of the effective range expansion for the three-dimensional
system are given in Table~\ref{3D_table}. \ One immediate application of our
three-dimensional results would be to calculate the reflection phase shift of
the deuteron from \textit{ab initio} lattice calculations using chiral
effective field theory. \ In addition to verifying the universal coefficients
in Table~\ref{3D_table}, it should be possible to measure the spin dependence
of the reflection phase shift resulting from the $D$-wave component of the
deuteron wavefunction.

\begin{table}[tb]
\caption{Coefficients of the effective range expansion for three-dimensional
dimer-wall scattering.}%
\label{3D_table}%
$%
\begin{tabular}
[c]{|c|c|c|c|}%
$m_{2}/m_{1}$ & $\kappa_{B}a_{R}$ & $\kappa_{B}r_{R}$ & $\kappa_{B}^{3}P_{R}%
$\\\hline\hline
1 & $0.279(8)$ & $-0.38(9)$ & $-0.02(2)$\\
2 & $0.342(8)$ & $-0.44(6)$ & $-0.02(2)$\\
4 & $0.506(8)$ & $-0.27(4)$ & $-0.02(1)$\\
8 & $0.731(8)$ & $0.03(2)$ & $-0.005(5)$%
\end{tabular}
\ \ \ \ \ \ $\end{table}

We now discuss the analytic calculation of the reflection scattering length
for a shallow dimer. \ The calculation we present uses the principles of
effective field theory but covers somewhat unfamiliar territory. Namely, our
system is inhomogeneous and there is no obvious small expansion parameter.

The method we use is to apply an adiabatic approximation for the
center-of-mass motion and generate an asymptotic expansion for the
long-distance part of the effective potential. \ We find that this process
generates an expansion for $\kappa_{B}a_{R}$ with convergence controlled by an
expansion parameter of size $e^{-2\kappa_{B}a_{R}}$. \ The derivation and full
details of this expansion up to second order will be given in a forthcoming
paper \cite{Lee:2011z}.\ We summarize the main method and results here.

Let the coordinates of the two constituent particles be $\vec{r}_{1}$ and
$\vec{r}_{2}$. \ The particles have a short-range attractive interaction, and
we take the Hamiltonian to be%
\begin{equation}
H=-\frac{1}{2m_{1}}\vec{\nabla}_{r_{1}}^{2}-\frac{1}{2m_{2}}\vec{\nabla
}_{r_{2}}^{2}+C_{B}\bar{\delta}^{(d)}(\vec{r}_{1}-\vec{r}_{2}),
\end{equation}
where $\bar{\delta}^{(d)}$ is a regulated $d$-dimensional delta function.
\ The coefficient $C_{B}$ is tuned to produce a bound state with energy
$E_{B}$ at infinite volume. \ Let $\vec{r}=\vec{r}_{1}-\vec{r}_{2}$ be the
relative separation between the particles. \ For any fixed center-of-mass
coordinate the dependence on the relative coordinate is given by%
\begin{equation}
H_{\text{rel}}=-\frac{1}{2\mu}\vec{\nabla}_{r}^{2}+C_{B}\bar{\delta}%
^{(d)}(\vec{r}).
\end{equation}

To calculate the reflection scattering length it suffices to compute
dimer-wall scattering in the limit $E_{K}\ll\left\vert E_{B}\right\vert $.
\ For this low-energy limit we use an adiabatic expansion for the
center-of-mass motion. \ This technique is conceptually similar to the
adiabatic hyperspherical approximation \cite{Macek:1968,Lin:1995a,Esry:1996a}.
\ For each $X$ we keep only the ground state for $H_{\text{rel}}$ satisfying
the hard-wall boundary condition. \ Contributions from higher excitations in
$H_{\text{rel}}$ are suppressed by powers of $E_{K}/\left\vert E_{B}%
\right\vert $. \ Therefore these higher excitations contribute to higher-order
coefficients in the effective range expansion but make no contribution to the
reflection scattering length.

For each $X$ let $\psi_{X}(\vec{r})$ be the normalized ground state
wavefunction of $H_{\text{rel}}$ satisfying the hard-wall boundary constraint.
\ The low-energy effective Hamiltonian is%
\begin{equation}
H_{\text{eff}}=-\frac{1}{2M}\frac{\partial^{2}}{\partial X^{2}}+V(X)+T(X),
\end{equation}
where $V(X)$ is the adiabatic potential,%
\begin{equation}
V(X)=\left\langle \psi_{X}\right\vert H_{\text{rel}}\left\vert \psi
_{X}\right\rangle ,
\end{equation}
and $T(X)$ is the diagonal adiabatic correction,%
\begin{equation}
T(X)=-\frac{1}{2M}\left\langle \psi_{X}\right\vert \frac{\partial^{2}%
}{\partial X^{2}}\left\vert \psi_{X}\right\rangle .
\end{equation}
We note that the cross-term involving the expectation value of $\psi_{X}$ for
one derivative with respect to $X$ vanishes due to the fixed normalization of
$\psi_{X}$.

Let us now define the length scales $r_{+}\left(  X\right)  =2MX/m_{1}$ and
$r_{-}\left(  X\right)  =2MX/m_{2}$. $r_{+(-)}\left(  X\right)  $ is twice the
distance from the wall to $\vec{r}_{1(2)}$ when the other particle touches the
wall. \ In other words $\psi_{X}(\vec{r})$ must vanish when the component of
$\vec{r}$ perpendicular to the wall equals $r_{+}(X)/2$ or $-r_{-}(X)/2$.
\ For large $X$ we can generate an asymptotic expansion for $V(X)$ and $T(X)$
in powers of $e^{-\kappa_{B}r_{+}\left(  X\right)  }$ and $e^{-\kappa_{B}%
r_{-}\left(  X\right)  }$.

At zeroth order we recover the infinite volume result, $V^{(0)}(X)+T^{(0)}%
(X)=E_{B}$. At first order the effective potential for the one-dimensional
case is%
\begin{align}
&  T^{(1)}(X)+V^{(1)}(X)\nonumber\\
&  = \frac{\kappa_{B}^{2}M^{2}}{m_{1}m_{2}}\left[  \frac{e^{-\kappa_{B}%
r_{+}\left(  X\right)  }}{m_{1}}+\frac{e^{-\kappa_{B}r_{-}\left(  X\right)  }%
}{m_{2}}\right]  . \label{V1_1D}%
\end{align}
In three dimensions the result is%
\begin{align}
&  T^{(1)}(X)+V^{(1)}(X)\nonumber\\
&  = \frac{\kappa_{B}M}{2m_{1}m_{2}X}\left[  e^{-\kappa_{B}r_{+}\left(
X\right)  }+e^{-\kappa_{B}r_{-}\left(  X\right)  }\right]  . \label{V1_3D}%
\end{align}
From these effective potentials it is straightforward to compute the
first-order result for the reflection scattering length. \ This process can be
carried forward to any order. \ The net result is an expansion with an
expansion parameter of size $e^{-\kappa_{B}r_{\pm}(a_{R})}$ $\le$
$e^{-2\kappa_{B}a_{R}}$. \ The larger $\kappa_{B}a_{R}$, the faster the
convergence of the expansion. \ First- and second-order results for the
one-dimensional system are shown in Fig.~\ref{latt_1D}, and results for the
three-dimensional system are shown in Fig.~\ref{latt_3D}. \ In each case the
agreement with lattice results is consistent with third-order corrections of
size $\le$ $e^{-6\kappa_{B}a_{R}}$.

We now finally address what happens in the limit $m_{2}/m_{1}\rightarrow
\infty$. \ Consider the limit $m_{2}\rightarrow\infty$ with $m_{1}$ held
fixed. \ In this limit the effective potential converges to a non-vanishing
finite-valued function, while the mass of the dimer grows with $m_{2}$. \ One
can check this explicitly for the expressions in Eq.~(\ref{V1_1D}) and
Eq.~(\ref{V1_3D}). \ Given the exponential tail of the effective potential,
the reflection radius near threshold has a logarithmic dependence on
$m_{2}/m_{1}$. \ An immediate application of these results would be to verify
the logarithmic $m_{2}/m_{1}$ threshold dependence in various physical
systems. \ This effect should be prominent for halo nuclei with a heavy core
and one satellite nucleon. It should also be seen in quantum dots and wells
for semiconductors with a large ratio between hole and electron effective
masses. It could also be reproduced with heterogeneous cold atomic dimers
consisting of one heavy and one light alkali atom.

In this letter we have discussed a number of phenomena relating to the elastic
scattering of quantum bound states from a hard surface. \ We have emphasized
processes of universal character which can be seen in several different
systems. \ The applications appear numerous, ranging from experimental
predictions for quantum dots and wells to numerical calculations of nuclear
structure. \ In our analysis we have introduced some theoretical techniques
which may be useful for describing the effective field theory of other
inhomogeneous systems.

\textit{The authors are grateful to Chris Greene, Hans-Werner Hammer, Ed
Marti, and Andreas Wieck for useful discussions. \ Financial support from the
U.S. Department of Energy and NC State GAANN Fellowship is acknowledged.
\ Computational resources provided by the J\"{u}lich Supercomputing Centre at
Forschungszentrum J\"{u}lich.}

\bibliographystyle{apsrev}
\bibliography{References}

\end{document}